%% file: main.tex
\pdfoutput=1
\documentclass[11pt]{article}

\usepackage[T1]{fontenc}
\usepackage[utf8]{inputenc}
\usepackage{lmodern}
\usepackage[a4paper,margin=1in]{geometry}
\usepackage{microtype}
\usepackage{booktabs}
\usepackage{tabularx}
\usepackage{array}
\usepackage{enumitem}
\usepackage{xcolor}
\usepackage{listings}
\usepackage{tikz}
\usetikzlibrary{positioning,arrows.meta,fit,backgrounds}
\usepackage[numbers,sort&compress]{natbib}
\usepackage{xurl}
\usepackage[hidelinks]{hyperref}
\usepackage{caption}

\setlist{nosep,leftmargin=1.6em}
\newcolumntype{L}{>{\raggedright\arraybackslash}X}
\newcolumntype{P}[1]{>{\raggedright\arraybackslash}p{#1}}
\newcommand{\code}[1]{\texttt{#1}}

\hypersetup{pdftitle={Record Sufficiency: What a Decision Record Must Preserve When an Unsettled Claim Becomes an Operative Status},
  pdfauthor={Amadeus Brandes}}

\input{results-macros}

\newif\ifcodingchecked
\ifnum\nCodingReviewed=\nStatements\relax\codingcheckedtrue\fi

\title{\textbf{Record Sufficiency}\\[4pt]
\large What a Decision Record Must Preserve When an Unsettled Claim\\ Becomes an Operative Status}
\author{Amadeus Brandes\\[2pt]
\normalsize Independent Researcher, Germany\\
\normalsize \texttt{brandesamadeus@gmail.com}}
\date{November 2026}

\begin{document}
\maketitle

\begin{abstract}
\noindent Organisations keep evidence, logs and approvals, but often not the step by which an unsettled claim became an operative status, such as an accepted label or a released image, so later readers cannot tell what the status was allowed to mean. We call a decision record \emph{sufficient} if it is true and a reader who has only the record and the items it cites can explain the status, bound it, re-derive it and challenge it. From these four uses we argue for ten questions a record must answer, map them onto the author's earlier JSON Schema for decision records (DRK v0.1), and define four mechanical checks. We separate conformance, which the checks decide; adequacy, whether the answers are specific enough to use, which needs a reader; fidelity, which no check on the record decides; and merits, which sufficiency does not require. On \nMutants{} single-defect variants of three synthetic records, schema validation, the only check DRK v0.1 implements, rejects every variant with a deleted required property and none of the others, and it accepts a record whose claim, rule, scope, limits and challenge route are empty. These are properties of the schema shown on defects we seeded, not estimates of performance on real records. A false record passes every check. The statement-of-reasons schema of the EU Digital Services Act Transparency Database answers \nSorFull{} of the questions in full, \nSorPartial{} in part and \nSorNone{} not at all: why the matter was unsettled, what the status depends on, whether it can be re-derived, and how to challenge it. In \nStatements{} real statements from \nStatementPlatforms{} platforms, drawn by a seeded rule and coded by a language model\ifcodingchecked{} and checked by the author\else{} whose grades the author has not yet checked\fi, free text addressed any of those four in only \nStatementsRaised, and in \nFiveNotWhy{} (\nFiveNotWhyStrict{} under a stricter reading adopted during coding) no free text said why the item fell under the rule. No deployment or legal claim is made.
\end{abstract}

\section{Introduction}

A dataset item receives an accepted label. An image is released for a campaign. A change in a clinical-trial registry is flagged as material. In each case somebody asserted that a thing should have a status, the matter was not obvious, and a person, a body or a rule settled it. Organisations usually keep the inputs to such decisions (votes, licences, registry versions, logs) and the outcome. They less often keep the step in between: which rule was applied, by whom, to which material, and with what limits. When that step is missing, a later reader cannot tell what the status was allowed to mean, and status words are read too broadly. ``Accepted'' is read as ``true'', ``released'' as ``cleared for any use'', ``material'' as ``misconduct''.

This paper is about the record of that step. A \emph{claim} is an assertion that something should have a status. A \emph{disposition} is the status that a competent authority or rule assigns in response.\footnote{``Disposition'' is used in its legal sense, the outcome by which a matter is disposed of. In records management the same word means the retention, transfer or destruction of records \citep{iso2016records}. That sense is not intended.} We call a claim \emph{unsettled} when neither inspection of the object nor the mechanical application of an agreed rule decides it: people disagree, the evidence is incomplete or conflicting, which rule governs or how it applies is a matter of judgment, or the answer depends on a condition that may change. A \emph{decision record} is the record of one disposition of one unsettled claim.

We ask what such a record must contain, and how much of that can be checked. The answer we propose is \emph{record sufficiency}: a decision record is sufficient when a reader who has only the record, and the items it cites, can explain why the status was assigned, bound what the status may be relied on for, re-derive the decision, and challenge it; and when what the record says about the decision is true. The term borrows from administrative law, where review is confined to the grounds that were given when the decision was taken \citep{chenery1943,regents2020}.

Three bodies of work come close. Reviewability asks what an organisation must keep so that automated decisions can be reviewed, and leaves record formats to future work \citep{cobbe2021reviewable}. Article 17 of the EU Digital Services Act prescribes a statement of reasons for one kind of decision, the restriction of content or accounts by a hosting provider \citep{eu2022dsa}. Work on AI agents published during 2026 proposes schemas for reasoning records and conditions under which claims about agents can be checked against records \citep{chang2026trace,zhao2026auditable,broemme2026evidence}. Design-rationale notations and architecture decision records keep the issues, options and arguments behind a decision \citep{kunz1970ibis,maclean1991qoc,nygard2011adr}, but not the authority, rule, scope, limits and route of challenge that an institutional status needs. None of these states, for an individual disposition in general, what the record must answer and why, and none measures how much of that a schema validator can see. Several of them draw a line between what a record can and cannot establish, and we build on that line.

The paper makes five contributions.
\begin{enumerate}
\item A definition of record sufficiency, and ten questions a decision record must answer, argued for from the four uses the record serves (Section~\ref{sec:rs}).
\item A separation of \emph{conformance}, \emph{adequacy}, \emph{fidelity} and \emph{merits}, which applies to decision records a distinction that law draws for reasons and that recent work draws for agent logs. Only conformance can be decided by generic mechanical checks; adequacy needs a reader, and fidelity needs something outside the record. Sufficiency requires adequacy and fidelity, and not merits (Section~\ref{sec:cfm}).
\item Four executable checks of conformance over the Disposition Record Kernel (DRK) v0.1, a JSON Schema with domain profiles that the author wrote earlier and onto which the ten questions are mapped (Section~\ref{sec:drk}).
\item An evaluation on synthetic records: a mutation study that illustrates what each check detects, the emptiest record the schema accepts, and two constructed records that mark the limit of every generic check (Section~\ref{sec:results}).
\item A test of the ten questions on a format designed independently of them: the statement-of-reasons schema of the Digital Services Act Transparency Database, and \nStatements{} real statements drawn from the database by a seeded rule (Sections~\ref{sec:dsa} and~\ref{sec:statements}).
\end{enumerate}

The claims are bounded. The case records are synthetic and written by the author. The only real records examined are \nStatements{} published statements of reasons, coded once, by a language model whose grades the author \ifcodingchecked checked\else has not yet checked\fi; there was no independent human coder. Nothing here shows that organisations would adopt such records, that they improve decisions, or that they satisfy any legal requirement. The individual fields are not new; Section~\ref{sec:related} identifies where each already appears. What the paper adds is the argument from use, the separation, and the measurements.

\section{Related work}\label{sec:related}

\subsection{Reasons, the record and timing in administrative law}\label{sec:law}

Administrative law has long treated the record of a decision as the object of review. In the United States, a court judges agency action on the grounds the agency itself gave: ``The grounds upon which an administrative order must be judged are those upon which the record discloses that its action was based'' \citep{chenery1943}. The ``focal point for judicial review should be the administrative record already in existence'', not a record assembled later in court; where the agency's explanation is too thin to permit review, the remedy is to obtain further explanation from the agency \citep{camp1973}. The Supreme Court restated the timing rule in 2020: review is limited to ``the grounds that the agency invoked when it took the action'', reasons supplied afterwards are suspect as ``post hoc rationalization'', and the Court gave as reasons that the rule promotes agency accountability and spares litigants and courts from chasing a moving target \citep{regents2020}. Later explanation is confined: where the original explanation indicates the determinative reason, the agency ``may elaborate later on that reason (or reasons) but may not provide new ones'' \citep{regents2020}. The administrative record in this sense is the body of material that was before the agency, closer to the evidence and logs on the left of Figure~\ref{fig:step} than to a record of one decision. What we take from it is the requirement that the grounds be legible from what existed when the decision was taken.

European Union law imposes a general duty to state reasons (Article 296 of the Treaty on the Functioning of the European Union; Article 41(2)(c) of the Charter of Fundamental Rights). The Court of Justice requires a statement of reasons to ``disclose in a clear and unequivocal fashion the reasoning followed by the institution'', and treats a missing or inadequate statement as a procedural defect that is distinct from the question whether the decision is lawful in substance \citep{sytraval1998}. Some legislatures fix the content of reasons in general terms. Under Australia's Acts Interpretation Act 1901, where an Act requires written reasons, the instrument giving them must ``set out the findings on material questions of fact'' and refer to the evidence or other material on which those findings were based \citep{cth1901aia}, a general rule that covers our Q3 and part of Q5. In English law, Cobbe distinguishes explanations of how a decision was made from reasons for why it was made, warns that public bodies may offer retrospective justifications, and notes that the law may need a way for a nominated decision-maker to show that they did not simply adopt the output of an automated system \citep{cobbe2019adminlaw}.

We take three ideas from this body of law. A record is to be judged by whether it permits review. The reasons that count are those relied on when the decision was taken. Whether reasons were adequately given is a different question from whether the decision was right. Sections~\ref{sec:def} and~\ref{sec:cfm} turn these into requirements on a record. The law motivates the design and does not entail it, and we make no claim that a record meeting our requirements meets any legal standard.

Outside the courtroom, Bovens defines accountability as ``a relationship between an actor and a forum, in which the actor has an obligation to explain and to justify his or her conduct, the forum can pose questions and pass judgement, and the actor may face consequences'' \citep{bovens2007}. The four uses in our definition correspond to what a forum does with an account: it seeks an explanation, it tests the limits of what is claimed, it checks the account, and it questions it.

\subsection{Reviewability, decision provenance and traceability}

Reviewability is the closest work in purpose. Cobbe, Lee and Singh decompose an automated decision-making process into technical and organisational stages (commissioning, model building, decision-making, investigation) in order to determine which record-keeping mechanisms are appropriate at each stage, so that individual decisions and the process as a whole can be reviewed \citep{cobbe2021reviewable}. Norval, Cobbe and Singh extend the idea to systems that span several organisations \citep{norval2021iot}. Decision provenance supplies the technical means, the tracing of data flows through interconnected systems \citep{singh2019decision}, and Kroll's account of traceability relates records of how a system was built and what it did to the goals of governance \citep{kroll2021traceability}.

The overlap with our problem is large: all of this work asks what must be retained so that decisions can be understood and questioned later. Three differences remain. First, reviewability is a method for deciding what to record, applied to a whole process and judged by whether the resulting information is relevant, accurate, proportionate and comprehensible in context. Its authors state that they do not prescribe how reviewable systems are to be built, and they name appropriate record-keeping formats as a subject for future guidance. We specify a format for one record, the record of one disposition, with content that can be enumerated and checked. Second, review under that framework generally proceeds by collating information recorded at several stages. We require the decision to be legible from one record made when the disposition was issued. Third, reviewability is scoped to automated decisions about natural or legal persons. Our cases concern objects (a dataset item, an image, a registry entry), although a person's interests can be implicated, as where the image shows an identifiable model; and only one of the three is decided by an automated process. Our proposal can be read as an answer, for one class of record, to a question reviewability leaves open. Closest in method is the same group's work on disclosure design, which treats information disclosures as interfaces designed for their recipients and finds in a user study that how a disclosure is structured affects whether it achieves its purpose \citep{norval2022disclosure}. Its object is the disclosure of how personal data are processed, not the record of a decision, but its result supports a premise of this paper: what a record contains and how it is organised matters, not only that information was kept.

Record-keeping has costs and side effects. A survey of 100 practitioners by Chappidi et al.\ reports that accountability-oriented record-keeping changes organisations; most respondents who reported a change described it as positive, but some described colleagues adjusting the timing and content of records, categorising events favourably, or falsifying records \citep{chappidi2025capture}. We return to this in Section~\ref{sec:discussion}; it is the empirical reason for separating fidelity from conformance.

\subsection{Statements of reasons}

Article 17 of the Digital Services Act obliges providers of hosting services to give affected recipients ``a clear and specific statement of reasons'' when they restrict content or accounts. The statement must state the restriction and, where relevant, its territorial scope and duration; the facts and circumstances relied on; whether automated means were used; the legal or contractual ground and why the content falls under it; and ``clear and user-friendly information'' on the possibilities for redress \citep{eu2022dsa}. Providers of online platforms also submit their statements to the Commission's Transparency Database, in a documented schema and without personal data \citep{ec2026dsadoc,ec2026dsaapi}. Empirical studies of the database report large differences between platforms in what they submit \citep{kaushal2024automated,trujillo2023dsa,drolsbach2023content}. Article 86 of the Artificial Intelligence Act gives persons affected by certain decisions based on high-risk AI systems a right to an explanation of ``the role of the AI system in the decision-making procedure and the main elements of the decision taken'' \citep{eu2024aiact}.

The statement of reasons is the most developed existing instance of a per-decision record with prescribed content. It was designed for one domain. Section~\ref{sec:dsa} maps its database schema to our ten questions.

\subsection{Justification and contestability}

Henin and Le M\'etayer argue that explainability does not suffice for the legitimacy of algorithmic decision systems, that justifiability and contestability are the key requirements, and that explanations and justifications differ in kind \citep{henin2022beyond}. Tajouri et al.\ develop the distinction: an explanation describes how a decision was made, a justification gives reasons that make it acceptable \citep{tajouri2026justification}. The distinction matches the legal one between how and why (Section~\ref{sec:law}), and we adopt it in the questions: the evidence concerns how the decision arose, the rule and the authority concern why the status holds. Work on contestability treats the ability to challenge a decision as a design property of systems \citep{alfrink2023contestable} and as a right \citep{kaminski2021contest}. A sufficient record is a precondition for contesting a decision, since one cannot contest a rule, an input or an authority that was never recorded. It does not create a right to contest or a procedure for doing so.

\subsection{Argument structure and reasoning records}

The ten questions resemble Toulmin's layout of an argument: a claim, the data offered for it, the warrant that licenses the step from data to claim, the backing of the warrant, a qualifier, and the conditions of rebuttal \citep{toulmin1958uses}. Our evidence is Toulmin's data, the rule is the warrant, the authority supplies the backing, the limits are the qualifier, and the conditions on which the status depends are the rebuttals. Assurance cases use similar structure to argue that a system has a property \citep{kelly2004gsn}; they are written by a developer about a system, not by a decider about a decision.

The closest work in form is TRACE, a typed and versioned schema for recording the reasoning of AI agents, with fields for the claim, its type, premises, warrants, evidence, defeaters, a verdict, failed gates and what is missing, and the discipline that no durable change of state occurs without a record \citep{chang2026trace}. Its authors state the limit that also governs our checks: gates verify the consistency of fields, not their truth. TRACE records the evaluation of an argument and ends in a verdict (accept, qualify, reject, defer). It treats authority as external to the record, and its schema lists no field for the period during which a status holds, for what the status does not establish, or for a route of challenge. We record a different object, the institutional disposition, and those are the fields that object needs.

\subsection{Design rationale and architecture decision records}

Several traditions record decisions together with their reasons. Issue-based information systems, devised for planning and policy deliberation, represent the issues raised, the positions taken and the arguments for and against them \citep{kunz1970ibis}. Design space analysis in human--computer interaction records the questions, options and criteria of a design \citep{maclean1991qoc}. Architecture decision records give each significant decision of a software team a short document with its context, the decision, its status (proposed, accepted, deprecated or superseded) and its consequences, and keep superseded records rather than deleting them \citep{nygard2011adr}. Architecture decision records are the closest practitioner format to a decision record, and DRK's field for superseded records serves the same purpose. None of these formats has a place for the authority or rule that settled the matter, the scope of the resulting status, what the status does not establish, or a route of challenge. They record deliberation, or a team's own choice that the team itself may revisit.

\subsection{Claims and evidence about AI agents}

Zhao et al.\ ask when a claim about an AI agent, such as ``a person approves every external email'', can be checked at all. Their answer is that the policy, the scope, the records that would settle the claim and their writers, and the decision rule must be fixed beforehand; a claim is then supported, contradicted or undecidable. They show that a record written only by the audited party cannot establish that nothing was omitted, and that integrity of a log is not completeness \citep{zhao2026auditable}. We rely on this result when we separate fidelity from conformance. Br\"omme models an evidentiary claim about an agentic process as a tuple that includes a \emph{limitation}, stating what the claim does not establish \citep{broemme2026evidence}. The author's own evidence records for changes in public registries list \code{claims\_supported} and \code{claims\_not\_supported} in each record \citep{brandes2026trialdiff}. An explicit statement of what a record does not establish is therefore not new with this paper. We require it of dispositions, and we derive the requirement from one of the four uses.

\subsection{Provenance and documentation}

Provenance models describe where an object came from and what was done to it \citep{moreau2013provdm}; content credentials do so for media \citep{c2pa2026spec}. Model cards and datasheets document models and datasets \citep{mitchell2019modelcards,gebru2021datasheets}. Decision modelling notations express the rule \citep{omg2024dmn}, and records-management standards govern how records are kept \citep{iso2016records}. A decision record cites all of these as evidence or as the rule applied. None of them is a record of a particular disposition.

\subsection{Summary}

Table~\ref{tab:comparators} places the closest work beside ours by the unit that is recorded. The comparison is by function, not by field names. The contribution claimed is narrow: an argument from use for the content a record must have, a separation of what can and cannot be checked, measurements of the first, and a mapping against one statutory format.

\begin{table}[t]
\caption{Closest work, by the unit recorded.}\label{tab:comparators}
\footnotesize
\begin{tabularx}{\linewidth}{@{}P{2.9cm}P{3.2cm}LL@{}}
\toprule
Work & Unit & What it fixes & What it leaves open \\
\midrule
Reviewability \citep{cobbe2021reviewable} & An automated decision process, by stage & A method for deciding what to record at each stage & Record formats and per-step content \\
DSA statement of reasons \citep{eu2022dsa} & One restriction of content or an account & Restriction, facts, ground, scope, duration, redress information & In the public database record: why unsettled, what the status depends on, re-derivation, route of challenge (Section~\ref{sec:dsa}) \\
TRACE \citep{chang2026trace} & One reasoning episode of an AI agent & Typed claim, evidence, warrants, defeaters, verdict & Authority; duration of a status; challenge \\
Auditable claims \citep{zhao2026auditable} & An operator's claim over many agent actions & Policy, scope, records and their writers, decision rule & Content of the individual decision record \\
Architecture decision records \citep{nygard2011adr} & One design decision of a software team & Context, decision, status, consequences; supersession & Authority and rule; limits; challenge \\
This paper & One disposition of one unsettled claim & Ten questions; four checks of conformance & Adequacy needs a reader; fidelity and merits cannot be checked from the record \\
\bottomrule
\end{tabularx}
\end{table}

\section{Record sufficiency}\label{sec:rs}

\subsection{Definition}\label{sec:def}

\begin{quote}
A decision record is \emph{sufficient} if (a) a reader who has only the record and the items it cites can explain why the status was assigned, bound what the status may be relied on for, re-derive the decision or see exactly why that is not possible, and challenge it; and (b) what the record states about the decision is true.
\end{quote}

Part (a) is a requirement of adequacy relative to four uses. All four are required. Replay may be satisfied by an explicit statement that the decision cannot be re-derived and why, since a reader who knows that a judgment was involved is better placed than one who assumes a computation. Part (b) is fidelity. It includes timing: the record states the grounds on which the decision was actually taken, which in practice means that the deciding authority makes the record when it issues the disposition. A record written later may be accurate, but a reader cannot tell it apart from a rationalisation (Section~\ref{sec:law}). When the record was made is a fact about the record, not about the decision, which is why none of the questions below asks it; DRK states it in a field beside the time the disposition was issued, and one of our checks tests the distance between them (relation R8, Section~\ref{sec:drk}).

The definition is of a property of a record. It says nothing about whether the decision was right.

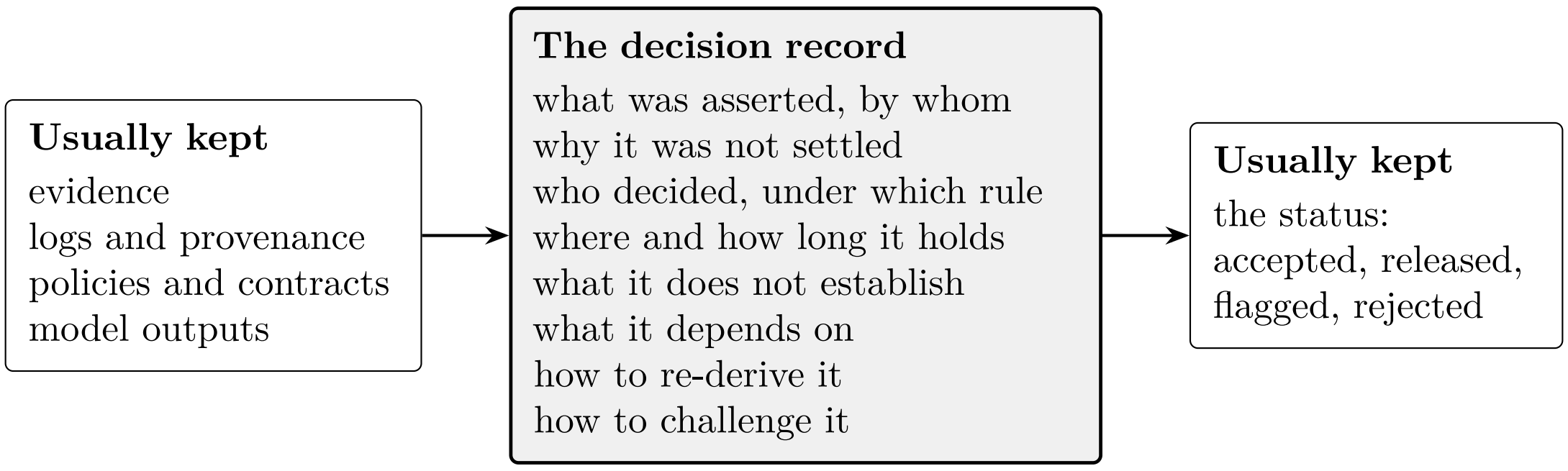
\begin{figure}[t]
\centering
\begin{tikzpicture}[font=\small,
  box/.style={draw,rounded corners=2pt,align=left,inner sep=6pt,text width=#1},
  arr/.style={-{Stealth[length=2.2mm]},thick}]
\node[box=3.4cm] (in) {\textbf{Usually kept}\\[2pt] evidence\\ logs and provenance\\ policies and contracts\\ model outputs};
\node[box=5.0cm,right=8mm of in,fill=black!6,line width=0.9pt] (mid) {\textbf{The decision record}\\[2pt] what was asserted, by whom\\ why it was not settled\\ who decided, under which rule\\ where and how long it holds\\ what it does not establish\\ what it depends on\\ how to re-derive it\\ how to challenge it};
\node[box=3.0cm,right=8mm of mid] (out) {\textbf{Usually kept}\\[2pt] the status:\\ accepted, released,\\ flagged, rejected};
\draw[arr] (in) -- (mid);
\draw[arr] (mid) -- (out);
\end{tikzpicture}
\caption{The step between evidence and status. Organisations usually keep what is on the left and on the right. The decision record preserves the step in the middle.}\label{fig:step}
\end{figure}

Figure~\ref{fig:step} shows where the record sits. It does not replace the evidence, which it cites, or the status, which it explains.

\subsection{Ten questions}\label{sec:questions}

What must a record contain to serve the four uses? We argue for ten questions (Table~\ref{tab:questions}). The argument for each is a removal test: take a record that answers the other nine questions and leaves this one unanswered, and ask which of the four uses a reader can no longer perform. A question earns its place if at least one use fails.

\emph{Explain.} A reader cannot explain a status without knowing what was asked (Q1), why a decision was needed at all (Q2), what material was considered (Q3), who or what decided (Q4), under which rule and how the rule was applied (Q5), and what resulted (Q6). Q2 also marks the boundary of the concept. If nothing was unsettled, there was no decision to record, and an ordinary event log is enough.

\emph{Bound.} To know what a status may be relied on for, a reader needs its scope and duration (Q6), an explicit statement of what it does not establish (Q7), and the conditions on which it depends (Q8). A recorded dissent or open question (Q2) also tells the reader how firmly the status may be relied on. Q7 is the answer to over-reading. Q8 states what must stay true for the status to hold, such as a licence remaining in force or a policy version remaining current, so that a reader can tell when the status has lapsed without anyone revoking it.

\emph{Re-derive.} To repeat a decision, a reader needs what was decided (Q1), the material (Q3), the rule and its application (Q5), the result to compare against (Q6), and a statement of whether repetition is possible and which elements, such as a reviewer's judgment, cannot be repeated (Q9).

\emph{Challenge.} A challenge needs a target and a route. The targets are the question that was decided (Q1), the material (Q3), the authority (Q4), the rule or its application (Q5), the scope of the status, since relying on it outside its scope is itself a ground of challenge (Q6), and the conditions that may have failed (Q8). The route is where a challenge goes and on what grounds it will be heard (Q10).

Each question is needed for at least one use, and three of them (Q7, Q9, Q10) for exactly one. We do not claim the list is the only possible decomposition. We claim that a record that fails to answer one of the ten fails one of the four uses, and that the ten can be put to any record format, including formats that were not designed with them in mind (Section~\ref{sec:dsa}). The questions were formulated with DRK v0.1 in view (Section~\ref{sec:drk}), so its having a section for each is not evidence for them.

\begin{table}[t]
\caption{The ten questions, the uses that need them (E explain, B bound, R re-derive, C challenge), and the DRK v0.1 section onto which the answer maps.}\label{tab:questions}
\footnotesize
\begin{tabularx}{\linewidth}{@{}lLccccP{4.3cm}@{}}
\toprule
 & Question & E & B & R & C & DRK v0.1 section \\
\midrule
Q1 & What was asserted, and by whom? & $\bullet$ & & $\bullet$ & $\bullet$ & \code{claim} \\
Q2 & Why was the claim not already settled? & $\bullet$ & $\bullet$ & & & \code{instability} \\
Q3 & On what material was it decided? & $\bullet$ & & $\bullet$ & $\bullet$ & \code{evidence\_inputs} \\
Q4 & Who or what decided? & $\bullet$ & & & $\bullet$ & \code{authority\_or\_rule}, \code{disposition} \\
Q5 & Under which rule, applied how? & $\bullet$ & & $\bullet$ & $\bullet$ & \code{authority\_or\_rule}, \code{disposition} \\
Q6 & What status resulted; where and how long does it hold? & $\bullet$ & $\bullet$ & $\bullet$ & $\bullet$ & \code{disposition}, \code{horizon\_or\_trigger} \\
Q7 & What does the status not establish? & & $\bullet$ & & & \code{limitations} \\
Q8 & What must stay true for the status to hold? & & $\bullet$ & & $\bullet$ & \code{support\_relation} \\
Q9 & Can the decision be re-derived, and what cannot be? & & & $\bullet$ & & \code{replay} \\
Q10 & On what grounds and where can it be challenged? & & & & $\bullet$ & \code{contestability} \\
\bottomrule
\end{tabularx}
\end{table}

\subsection{Conformance, adequacy, fidelity and merits}\label{sec:cfm}

Four properties of a decision record are easily run together.

\emph{Conformance} is what generic mechanical checks can decide from the record and its profile alone: that the record has the required structure, that it carries the fields its domain requires, that the fields answering the ten questions are not empty, and that its parts agree in ways a check can see without knowing what the rule means. Section~\ref{sec:drk} defines four such checks for DRK.

\emph{Adequacy} is part (a) of the definition: the answers are specific enough for a reader to explain, bound, re-derive and challenge the decision. Most conformance conditions are necessary for adequacy. A record that fails the content check leaves an answer out, one that fails the profile check lacks what its domain requires, and one that fails a coherence relation other than R8 contradicts itself. Two kinds of condition serve other ends: the timing relation R8 tests when the record was made, which belongs to fidelity, and the identifying fields the schema requires let a reader cite the record rather than use it. No conformance condition is sufficient for adequacy. A record that passes every check can still answer with a formula, such as a route of challenge that would fit any decision, and it can omit evidence that its own rule needs (Section~\ref{sec:shipped}); a reader finds the second only by reading the rule. Adequacy is judged by a reader.

\emph{Fidelity} is part (b), the truth of what the record says about the decision: the evidence listed was the evidence considered, the named authority decided, the rule cited was the rule applied, and the record was made when it says it was. Fidelity cannot be decided from the record alone. Any procedure that reads only the record gives the same answer for a true record and for a false record of the same form; Zhao et al.\ prove the corresponding statement for claims about agents \citep{zhao2026auditable}. Establishing fidelity takes something outside the record: inspection of the cited items, or a second record written by a party the decider does not control.

\emph{Merits} are whether the decision was right: whether the rule was appropriate, whether it was applied correctly, whether the evidence was enough to support it. Sufficiency does not require a correct decision. A sufficient record of a wrong decision is what allows a reviewer to find the error, which is the distinction European law draws between the adequacy of a statement of reasons and the legality of the decision \citep{sytraval1998}.

\begin{figure}[t]
\centering
\begin{tikzpicture}[font=\small,
  txt/.style={align=left,inner sep=0pt,text width=7.0cm},
  note/.style={align=left,text width=4.1cm,font=\footnotesize}]
\node[draw,fill=black!16,align=left,inner sep=4pt,text width=7.0cm] (c) {\textbf{Conformance}\quad structure, profile, content, coherence};
\node[txt,above=2.4mm of c] (at) {\textbf{Adequacy}\quad the answers are specific enough to use};
\begin{scope}[on background layer]
\node[draw,fill=black!8,inner xsep=3mm,inner ysep=2.4mm,fit=(c)(at)] (a) {};
\end{scope}
\node[txt,above=4.8mm of at] (ft) {\textbf{Fidelity}\quad the record reports what happened};
\begin{scope}[on background layer]
\node[draw,fill=black!3,inner xsep=0pt,inner ysep=2.4mm,fit=(ft)(a.west|-ft)(a.east|-ft)] (f) {};
\end{scope}
\node[txt,above=7.8mm of ft] (mt) {\textbf{Merits}\quad the decision was right};
\node[draw,dashed,inner xsep=0pt,inner ysep=2.4mm,fit=(mt)(a.west|-mt)(a.east|-mt)] (m) {};
\node[note,anchor=west] at ([xshift=4mm]a.east|-c) {decided from the record alone by checks C1 to C4};
\node[note,anchor=west] at ([xshift=4mm]a.east|-at) {judged by a reader of the record};
\node[note,anchor=west] at ([xshift=4mm]f.east) {needs the cited items or an independent record};
\node[note,anchor=west] at ([xshift=4mm]m.east) {needs review of the decision; not part of sufficiency};
\draw[thick] ([xshift=-3mm]a.south west) -- node[left=2mm,align=right,font=\footnotesize,text width=1.7cm] {record\\ sufficiency} ([xshift=-3mm]f.north west);
\draw[thick] ([xshift=-3mm]a.south west) -- ++(1.2mm,0);
\draw[thick] ([xshift=-3mm]f.north west) -- ++(1.2mm,0);
\end{tikzpicture}
\caption{Four properties of a decision record. Conformance consists of conditions that generic checks decide; most of them are part of adequacy, and the timing relation R8 bears on fidelity. Sufficiency is adequacy together with fidelity. Merits lie outside it.}\label{fig:layers}
\end{figure}
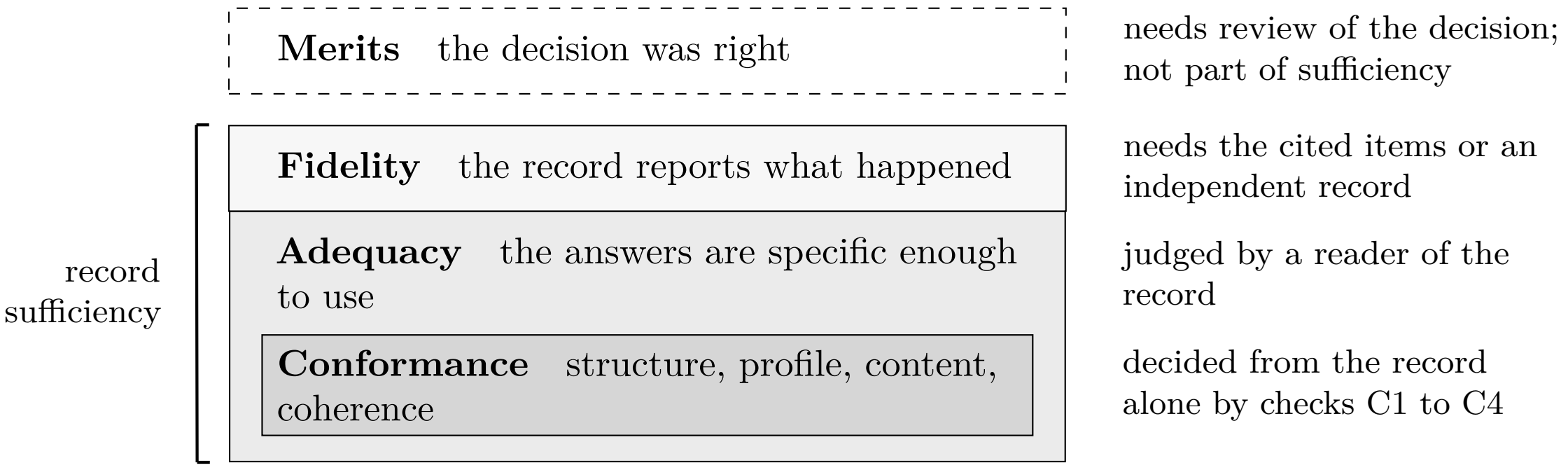

Sufficiency is adequacy together with fidelity (Figure~\ref{fig:layers}). Conformance is largely necessary for adequacy and never sufficient for it, and a sufficient record need not record a correct decision. Section~\ref{sec:limits} constructs a conformant and adequate record that is false, and a sufficient record of a wrong decision. The documentation of DRK v0.1 names three validation levels: kernel shape, profile completeness and domain validity. The first is our C1. The second, which the documentation recommends for v0.2, is the core of our C2. The third combines what we separate as adequacy, fidelity and merits.

\subsection{Terms}

The paper introduces few terms. \emph{Record sufficiency} adapts a phrase from administrative law. \emph{Adequacy} echoes the law's question whether reasons were adequately stated (Section~\ref{sec:law}) but is defined here by the four uses. \emph{Disposition}, \emph{authority}, \emph{rule}, \emph{scope} and \emph{challenge} are used in their ordinary legal senses. \emph{Non-claims} is our short name for the answer to Q7, what a status does not establish. The Disposition Record Kernel uses two names that need a gloss. Its \code{instability} section answers Q2, why the claim was not settled. Its \code{support\_relation} section answers Q8: it lists what the status depends on, who maintains those conditions, and what would undo them. The name comes from an earlier working paper by the author on claims that cease to hold when the party maintaining them withdraws \citep{brandes2026state}.

\section{The Disposition Record Kernel and four checks}\label{sec:drk}

\subsection{DRK v0.1}

The Disposition Record Kernel (DRK) v0.1 is a JSON Schema \citep{jsonschema2020} for decision records, with five domain profiles and five example records. It was produced by the author in June 2026 and is distributed with this paper without modification. Its documentation already pairs each of its main sections with a question. The ten questions of Section~\ref{sec:questions} were formulated later, with DRK in view; Table~\ref{tab:questions} maps them onto its sections, and DRK does not implement them in any stronger sense. A record has \nRequiredProps{} required top-level properties: five that identify it and twelve sections, of which ten carry the answers to the ten questions (Table~\ref{tab:questions}). The remaining two are \code{provenance}, which states how the record itself was produced, and \code{profile\_extensions}, which holds domain-specific fields. The schema fixes vocabularies for the kind of instability (\nInstabilityTypes{} values), authority (\nAuthorityTypes), horizon (\nHorizonTypes) and evidence (\nEvidenceTypes). A profile names the extension fields its domain requires (\nProfileExtMin{} or \nProfileExtMax{} per profile) and recommends disposition values. The \nProfiles{} profiles cover label resolution, visual release, clinical-trial materiality, custody admission of software artefacts, and withdrawal of support for a claimed state. This paper uses the first three.

The schema is deliberately permissive about content. It requires every section to be present and typed, but it sets a minimum length for only two lists (the instability types and the evidence inputs) and constrains no free-text field. The schema therefore decides whether a record has the shape of a decision record. It does not decide whether the record answers the ten questions. The schema's documentation states that v0.1 does not fully validate profile-specific fields.

\subsection{Four checks}

We define four checks of conformance. C1 is the validation that DRK v0.1 implements. C2 tests the profiles' lists of required fields, which DRK's documentation recommends checking from v0.2 on; its tests that a required field is not empty and that the claim classes agree are ours. C3 and C4 are defined in this paper. Appendix~\ref{app:checks} lists all four in full.

\begin{description}[leftmargin=1.6em,style=nextline]
\item[C1 Structure.] The record validates against the DRK v0.1 schema, with date-time formats enforced.
\item[C2 Profile.] The record names a profile, its claim class matches the profile, and every extension field the profile requires is present and not empty.
\item[C3 Content.] Each of \nContentFields{} fields that carry the answers to the ten questions is not empty. If the record says the decision can be re-derived, it names the inputs and gives instructions; if it says it cannot, it names what cannot be repeated.
\item[C4 Coherence.] Nine relations between fields hold. The evidence named as the rule's required inputs or as the replay's inputs is listed in the record; evidence identifiers are unique; the issuer of the disposition is the named authority; no evidence and no claim is dated after the disposition; the horizon does not end before it begins; the record was created no earlier than the disposition and no more than a set interval after it (24 hours by default); and the record does not name itself as its own predecessor.
\end{description}

Relation R8 is the mechanical part of the timing requirement. It flags a record written long after the decision it reports. It cannot prove that a record was written when it says, because a timestamp is a statement in the record like any other.

C3 could largely be expressed as additional constraints in the schema. Most of C4 cannot, because JSON Schema does not express references between parts of a document or comparisons between dates. The checks are implemented in Python and tested by \nSelfTests{} self-tests.

No check can establish fidelity or merits. Nor can a generic check establish adequacy. It cannot distinguish a specific answer from a formula: a field that says ``route to the responsible reviewer'' is not empty. And it cannot tell whether the listed evidence includes what the stated rule needs, which requires reading the rule.

\section{Method}\label{sec:method}

The work follows the design-science pattern of building an artefact for a stated problem and evaluating it \citep{hevner2004design,peffers2007dsrm}. The artefact is the definition, the ten questions and the four checks. The evaluation is analytical and experimental, on synthetic records and on one sample of real records. It addresses five questions.

\begin{description}[leftmargin=1.6em,style=nextline]
\item[EQ1] Do the checks accept records written to be sufficient?
\item[EQ2] What does each check detect, and what does schema validation alone miss?
\item[EQ3] What can no generic check detect?
\item[EQ4] Do the ten questions discriminate when applied to a record format that was designed independently of them?
\item[EQ5] Do real records in that format answer what the format can carry, and does their free text supply what it cannot?
\end{description}

\emph{Materials.} We use DRK v0.1 unchanged, with its five example records. We wrote three case records for this paper, one for each of three profiles (Section~\ref{sec:cases}), and two variants of them (Section~\ref{sec:limits}). All are synthetic. They document no real decision, person, product or trial. For EQ4 we use the submission schema of the DSA Transparency Database as documented by the European Commission on 16 November 2026 \citep{ec2026dsaapi}. For EQ5 we use \nStatements{} statements of reasons published in that database (Section~\ref{sec:statements}).

\emph{Mutation study.} For EQ2 we use mutation analysis, a standard way to measure what a test detects by seeding defects and counting how many are found \citep{demillo1978hints}. From each case record we derive mutants that differ from it in exactly one respect, in four classes: (A) one required top-level property deleted; (B) one of the fields that answer the ten questions emptied in a way the schema's types allow; (C) one profile-required field deleted; (D) one cross-field relation broken, for example a rule that cites evidence the record does not list, a record dated ninety days after the disposition, or a record that names itself as superseded. This gives \nMutants{} mutants (\nClassA, \nClassB, \nClassC{} and \nClassD{} per class). We run all four checks on every mutant. Because the author designed both the checks and the classes, the study illustrates what the schema and the checks can see; it does not measure them.

\emph{Mapping.} For EQ4 we assign each of the \nSorAttr{} attributes of the statement-of-reasons schema to the question it helps answer, and grade each question: \emph{full} if an attribute dedicated to the answer carries it, whether as a structured value or as free text; \emph{partial} if the answer can only be inferred from attributes that serve another purpose, or if a dedicated attribute can carry only part of it; \emph{none} if no attribute carries it. We count EQ4 as failed if a question cannot be graded from the attribute list, or if every question receives the same grade. We then convert an illustrative statement, written by us with the documented attribute names, into a DRK record and run the checks on it.

\emph{Sample of real statements.} For EQ5 the sampling and coding rules were written down before any sampled statement was opened, and amended twice during sampling when the database returned errors; the protocol and its amendments are in the ancillary files. The times of the amendments were reconstructed afterwards from file modification times. The \nPlatformsListed{} platforms in the database's platform filter were put in an order fixed by a seeded shuffle. For each platform in turn we took the first statement the database's search page lists for 2 November 2026, skipping platforms with no statement that day, until \nStatements{} statements had been drawn; \nPlatformsTried{} platforms were tried. For \nDayQueryFailed{} platforms the database returned an error to the query filtered by day; under the first amendment we took the platform's most recent statement instead, which added \nStatementFallback{} statements received on \fallbackDays, so that \nStatementsSampleDay{} statements come from 2 November. Under the second amendment, a platform whose statement page could not be retrieved was skipped; this happened for \nStatementUnavailable{} platform. Each statement is graded on each question with the grades used for the schema, applied to its content. Under the protocol, text is \emph{generic}, and earns at most \emph{partial}, if it ``would be equally true of any decision of the same type and ground on that platform''. How to treat text that names a reason narrower than the stated ground, as a finding about the item, was not fixed in advance. After the first 13 statements had been read, and before any grade was recorded, the coding adopted a stricter reading under which such text is generic. Because the rule as written came first, we report its grades as primary and the stricter reading beside them. Under the rule as written, text earns full for Q1, Q3 and Q5 if, beyond the stated ground and the category, it states what feature or behaviour of the item the decision rests on, even when it is phrased as the condition of a narrower rule; naming a narrower rule document or provision without such a feature earns partial. Under the stricter reading only a statement of fact about this item earns full. The coding reads every free-text field of a statement, not only the explanation. The grades were produced by a language model under the protocol's rules and revised once, after an independent review by a separate language-model agent, chiefly because the stricter reading had been reported as if it were the rule. \ifcodingchecked The author then checked them, as the protocol requires.\else They have not yet been checked by the author, as the protocol requires before submission.\fi Two schema-level grades were corrected before any statement was coded, and a third after coding (Section~\ref{sec:dsa}). A keyword screen on the free-text fields, which needs no judgment, is run beside the coding for Q2, Q7, Q8 and Q10 as a partial check on it. It does not cover Q1, Q3 and Q5, where the two readings differ.

\emph{Reproducibility.} The evaluation is deterministic. One command regenerates every number in this paper that comes from the evaluation, and the bodies of Tables~\ref{tab:mutation}, \ref{tab:dsa} and~\ref{tab:statements}, from the schema, the profiles, the records, the saved statement pages and the script that holds the coding, and stops with an error if any of them differs from the copy compiled into the paper. It also checks, against a saved copy of the database's search page, that the platforms tried and sampled follow the seeded order. Drawing the sample needs network access and is not part of that command; the statements as retrieved are included.

\emph{Threats to validity.} The author wrote the schema, the records, the checks and the mutation classes. Checks C2 to C4 detect the mutants of classes B to D by construction, and classes B to D were designed to remain valid under the schema, so the pattern of schema validation is by construction too. The number of mutants in each class was also chosen by the author, so a detection rate pooled over classes describes the design, not the checks; we report results per class. What carries evidential weight are existence results: a record the schema accepts that answers almost nothing, shipped examples that pass every check and are not adequate, and two records that defeat every generic check. The mapping in EQ4 was made by one person, and the questions were not formulated blind to the DSA schema, which the author had examined earlier, or to DRK. It is a test of discrimination, not of a prediction. The coding in EQ5 has not been repeated by an independent coder, and whether a text is generic is a judgment. Synthetic records show that the format can express a decision. They do not show how real records would look.

\section{Three cases}\label{sec:cases}

Table~\ref{tab:cases} summarises the three case records. Each concerns a status that is commonly read too broadly.

\begin{table}[t]
\caption{The three synthetic case records (abridged).}\label{tab:cases}
\footnotesize
\begin{tabularx}{\linewidth}{@{}P{1.9cm}LLL@{}}
\toprule
 & Label resolution & Visual release & Trial registry change \\
\midrule
Claim (Q1) & Item should carry the label \code{medical\_device} in a dataset release & Image may be published for a campaign, on the owned landing page and in paid social media & Change between two registry versions should be treated as material for review \\
Why unsettled (Q2) & Annotators split 2:1; a model prelabel sides with the minority; the boundary is set by policy & Depends on the scope of the licence and of the model's consent & A text difference does not show whether a change matters \\
Evidence (Q3) & Item text, three votes, prelabel, policy, adjudicator note & Origin record, content credential, licence, model release, policy, reviewer note & Two registry versions, their difference, recorded study start date \\
Authority and rule (Q4, Q5) & Adjudication reviewer; label policy v0.4 & Release review; release policy v0.2 & Deterministic rule: a change to a primary outcome after enrolment began is material unless administrative \\
Disposition and scope (Q6) & Label accepted, for this release under this policy & Released with restrictions: landing page only, three countries, six months, unaltered & Material for review, for this version comparison only \\
Does not establish (Q7) & That the product is a regulated medical device; that the label is correct apart from the policy & Clearance for paid social media, edits, other territories or later dates; that the credential confers a right to publish & Misconduct, bias, harm, legal breach, or that the outcome was not prespecified in the protocol \\
Depends on (Q8) & Policy version in force; item text unchanged & Licence and consent remaining valid & Registry versions and start date being correct \\
Challenge (Q10) & Re-adjudication by a different reviewer & Rights desk for rights holders; re-review ticket internally & Sponsor may submit protocol evidence; a reviewer re-applies the rule \\
\bottomrule
\end{tabularx}
\end{table}

\emph{Label resolution.} Disagreement between annotators is often genuine and carries information \citep{aroyo2015truth,plank2022problem}. A dataset release still needs one operative label. The record keeps the two apart. It states the dissenting vote and the dissenting prelabel, names the policy criterion that decided the matter, treats the model output as evidence and not as a decision, and says that the label is the result of applying a policy to a text. A later user who trains on the label can see that it is not a finding about the product.

\emph{Visual release.} A content credential records the origin and edit history of a file \citep{c2pa2026spec}. It says nothing about who may publish it or where. In the record, the request is broader than what is granted: the campaign team asks for the landing page and paid social media, and the review grants the first and excludes the second because the licence does not cover it. The horizon has an end date, the day the licence term ends, and the dependence field states that the release lapses if the licence or the model's consent is withdrawn.

\emph{Trial registry change.} Changes to registered outcomes are a documented concern. Mathieu et al.\ found discrepancies between registered and published primary outcomes in 31\% of adequately registered trials \citep{mathieu2009comparison}, and the COMPare project found outcome discrepancies requiring correction in 58 of 67 trials in five leading journals \citep{goldacre2019compare}. Whether a particular change matters is harder to settle. In earlier work on evidence records for registry changes, the author found that membership of a change in a defined class could be verified against the registry, while reviewers asked to confirm a priority label did not converge \citep{brandes2026trialdiff}. The disposition in this case is accordingly the output of a stated rule. The claim is unsettled all the same: which changes count as material is the review team's choice of rule, which a sponsor may dispute; the rule's exception for clearly administrative changes calls for judgment; and whether the new outcome was prespecified elsewhere is an open question that the registry cannot answer. The rule settles the status for review triage; it does not make the claim self-evident. The record says what the rule is, lists the recorded study start date as evidence because the rule depends on it, and states that the flag establishes neither misconduct nor that the change was not prespecified in a protocol.

Listing~\ref{lst:visual} shows part of the visual-release record.

\begin{lstlisting}[caption={Excerpt from the visual-release case record (lists abridged).},label={lst:visual},float=t]
"disposition": {
  "disposition_value": "released_with_restrictions",
  "rationale": "The licence and the model release cover owned-website use of the
     unaltered asset in three territories for six months. Neither covers paid
     social or edited versions, so the request is granted in part.",
  "issued_by": "Visual Release Review",
  "issued_at": "2026-06-11T12:00:00Z",
  "effective_scope": "Owned campaign landing page; unaltered asset; DE, AT and CH;
     2026-06-11 to 2026-12-11." },
"limitations": {
  "claims_not_supported": [
    "That the asset is cleared for paid social, out-of-home or third-party use.",
    "That the content credential establishes a right to publish." ] },
"contestability": {
  "challenge_conditions": [
    "The model release does not cover this use, or consent has been withdrawn." ],
  "review_path": "The licensor and the depicted model may write to the rights desk
     named in the licence. Internal challengers file a re-review ticket with
     Visual Release Review. Publication is suspended while a challenge from a
     rights holder is open." }
\end{lstlisting}

\section{Results}\label{sec:results}

\subsection{The checks accept the case records and the shipped examples}\label{sec:shipped}

The three case records pass all four checks. So do the \nExamples{} example records distributed with DRK v0.1, and check C1 reproduces the validation result recorded in that package.

That the shipped examples pass is informative in an unwelcome way, because on reading they are not adequate: they are conformant, but their answers are not specific enough to use. They were generated from a template. All \nContestIdentical{} of the \nContestLeaves{} text values in their \code{contestability} sections are identical across the five domains, so a label dispute and a software admission share one generic route of challenge. The trial example applies a rule that depends on when enrolment began and lists no evidence of that date. The visual example describes a six-month term and gives its horizon no end date. Each of these is a defect of substance in a record that is well formed. A reader finds them; no generic check that reads one record can. The case records written for this paper correct them. The observation that records turn into template text is a known risk of record-keeping; here it occurred in the artefact's own examples.

\subsection{The emptiest record the schema accepts}

How little can a record say and still validate? We reduced each case record to the emptiest value the schema permits at every position: empty strings where no pattern, format or vocabulary applies, empty lists where no minimum length is set, and null where null is allowed. The result passes C1. In it, all \nContentEmpty{} of the \nContentFields{} fields that answer the ten questions are empty, and \nBlankLeaves{} of its \nStringLeaves{} string values are blank; the others are identifiers, dates and vocabulary terms that the schema constrains. It names no claim, no claimant, no rule, no rationale, no scope, no non-claims and no route of challenge. Checks C2 and C3 reject it. A record can therefore be valid under the v0.1 schema and answer none of the ten questions beyond what fixed vocabularies and flags say: the kind of instability, authority, evidence and horizon, that the status is active, and that the decision can be re-derived, without saying how.

\subsection{Mutation study}\label{sec:mutation}

Table~\ref{tab:mutation} gives the number of mutants each check rejects. Schema validation rejects every mutant with a deleted required property and no other mutant. It misses every emptied field, every missing profile field and every broken relation, \nMissedByCone{} mutants in all. The pattern is all or nothing by class, and it follows from the schema and from how the classes were built: the schema requires each property to be present, sets no minimum content for free-text fields, does not validate profile extensions, and cannot express relations between fields, and classes B to D stay within what it accepts. Pooled over the four classes, schema validation rejects \nMutCone{} of the author's \nMutants{} seeded mutants. That figure is the share of class-A mutants in our design and nothing else; with a different mix of classes it would be different. It is not an estimate of how often schema validation catches defects in real records, whose defects we have not sampled. Each of the other three checks detects the class it was written for, as it must. C4 also rejects \nBcaughtByCfour{} class-B mutants, because emptying the name of the authority or of the issuer breaks the relation between them.

\begin{table}[t]
\caption{Mutants rejected by each check. Three case records; each mutant differs from its original in one respect.}\label{tab:mutation}
\small
\begin{tabularx}{\linewidth}{@{}Lrrrrrr@{}}
\toprule
Mutation class & Mutants & C1 & C2 & C3 & C4 & Any \\
\midrule
\input{tab-mutation.tex} 
\bottomrule
\end{tabularx}
\end{table}

Read by class, the C1 column shows what the schema's definition already implies: the properties that distinguish a decision record from a log entry (the non-claims, the route of challenge, the scope, the link between a rule and its evidence) are those the schema does not enforce. The study illustrates this; it does not discover it.

\subsection{What no generic check can see}\label{sec:limits}

Two constructed records mark the limit of the four checks.

\emph{A false record.} The first variant reports the label decision of Section~\ref{sec:cases} untruthfully. The dissenting vote and the dissenting prelabel are rewritten as agreement, the statement of residual uncertainty is removed, and the strength of the decision is given as unanimous. It differs from the true record in \nFalseDiff{} values, counting its identifier. It passes all four checks, and on reading it is adequate. Nothing in the record betrays the change, and nothing could: fidelity is a relation between the record and the event, and the checks see only the record. The falsification would be found by comparing the record with the annotation platform's own log, that is, with a record the adjudicator did not write.

\emph{A misapplied rule.} The second variant is truthful and complete. The deciding system applied the trial rule to a change made before the study started. The evidence gives an anticipated start date later than the change; the rule requires the change to follow the start of enrolment; the record's account of the decision method says that the system read the start date but did not compare it with the date of the change; and the disposition is nevertheless ``material''. The variant differs from the correct record in \nMisappliedDiff{} values, counting its identifier, and passes all four checks. Here the error is legible to a reader. Re-deriving the decision as the record instructs (Q9), by applying the stated rule (Q5) to the listed evidence (Q3), gives a different status, and the record says where to send the objection (Q10). No generic check finds the error, because finding it requires applying the rule; a check written for this one rule could. This is the case the definition is meant to permit. The record is sufficient, the decision is wrong, and the record is what makes the error findable.

The two variants mark the two limits of conformance. A conformant and adequate record may be false, and then only something outside the record can show it. A sufficient record may document a bad decision, and then the record itself is the means of showing it.

\subsection{The DSA statement of reasons}\label{sec:dsa}

The three cases were written by the author to fit the questions. To see whether the questions discriminate on a format that was not, we apply them to the statement of reasons as submitted to the DSA Transparency Database. The submission schema, as updated on 1 July 2025, has \nSorAttr{} attributes \citep{ec2026dsaapi}. This subsection grades the schema, that is, what the format can carry. Section~\ref{sec:statements} grades real statements.

\begin{table}[t]
\caption{The ten questions applied to the statement-of-reasons schema of the DSA Transparency Database.}\label{tab:dsa}
\footnotesize
\begin{tabularx}{\linewidth}{@{}P{5.6cm}lL@{}}
\toprule
Question & Coverage & Attributes \\
\midrule
\input{tab-dsa.tex} 
\bottomrule
\end{tabularx}
\end{table}

All \nSorAttr{} attributes map to a question (Table~\ref{tab:dsa}). The schema answers \nSorFull{} questions in full. Every published statement names the deciding platform, a field the database adds to what is submitted, and says whether the decision was automated. A required free-text attribute is dedicated to the facts and circumstances relied on, beside a flag for automated detection. The schema states the rule precisely: the legal or contractual ground, an explanation of why the content falls under it, and a reference. It states the status, its territory, its start date and an end date for each kind of restriction. It answers \nSorPartial{} questions in part. The affected content and the source of the notice are identified, but the assertion is not recorded apart from the ground and the restriction. For a restriction on contractual grounds, an optional yes-or-no attribute can state that the content is not considered illegal; that is one fixed limit, and the only one the format can express. For \nSorNone{} questions there is no attribute: why the matter was not settled, what the restriction depends on, whether it can be re-derived, and how to challenge it.

Converting an illustrative statement into a DRK record makes the gap concrete but adds no evidence. The \nSorMissingSections{} sections that would carry those answers have no source in the statement, so the record lacks them and fails C1, and C3 reports \nSorContentFindings{} empty answers. The statement is one we wrote with the documented attribute names. The record's identifying fields, and the fields DRK fills from fixed vocabularies (the types of authority, horizon, evidence and disposition, the status and the profile), are set to constants, and C2 is not run, since no profile exists for statements of reasons. Its section on limits exists only because the statement answers the attribute on illegality with no, which \stSevenNone{} of the \nStatements{} real statements in Section~\ref{sec:statements} do not. C4 also fails, on relation R8, because the schema gives the date of application without a time and we read it as midnight.

Three grades in this table were corrected. Before the coding of real statements began, Q4 was raised from partial to full: grading it partial because the deciding role or body is not named applied a stricter standard than DRK applies to its own authority field, which a platform's name satisfies. Q7 was raised from none to partial, because the yes-or-no attribute on illegality had been overlooked. After coding, Q3 was raised from partial to full: the earlier grade, for want of itemised evidence, applied a stricter standard than the one used for Q4 and than DRK's own evidence section, which asks only for a summary and a source per item. Q5 is full because its explanation attribute is dedicated to why the content falls under the ground, although it is free text.

The absence of a challenge route needs care. Article 17(3)(f) requires redress information in the statement sent to the affected recipient. The database schema has no attribute for it, and the Commission's documentation of the datapoints refers to points (a) to (e) of Article 17(3) and not to point (f) \citep{ec2026dsadoc}. The regime grants the remedy; the public record does not carry the route. That is not a defect of the database, whose purpose is transparency at scale and which must exclude personal data. It shows that the same decision can have one record that carries the route of challenge, the statement sent to the recipient, and a public record that does not. We did not examine the statements as sent. The ten questions make the difference visible.

The mapping supports two limited conclusions. The questions discriminate by the criterion of Section~\ref{sec:method}: every question could be graded, and the grades vary. That the format is strong where a transparency instrument would be expected to be strong (who decided, the material, the rule, the status and its scope) is an explanation offered after the grading, not a test. And the questions it leaves open touch all four uses: why the matter was unsettled (Q2) bears on explaining and bounding, what the status depends on (Q8) on bounding and challenging, re-derivation (Q9) on re-deriving, and the route (Q10) on challenging.

\subsection{Real statements of reasons}\label{sec:statements}

A schema says what a record can carry. Free-text fields can carry more, and filled fields can say less than the schema provides for. Table~\ref{tab:statements} gives the grades for \nStatements{} real statements from \nStatementPlatforms{} platforms, drawn as described in Section~\ref{sec:method}. The platforms range from large international services (Reddit, Pinterest, Airbnb, Twitch, Chrome Web Store) to national marketplaces and a news site.

\begin{table}[t]
\caption{The ten questions applied to \nStatements{} real statements of reasons, one per platform, coded by a language model\ifcodingchecked{} and checked by the author\else{} and not yet checked by the author\fi{} (Section~\ref{sec:method}). \emph{Schema} repeats the grade from Table~\ref{tab:dsa}; the other columns count statements under the coding rule as written, and the last column counts full grades under the stricter reading adopted during coding.}\label{tab:statements}
\footnotesize
\begin{tabularx}{\linewidth}{@{}Llrrrr@{}}
\toprule
Question & Schema & Full & Partial & None & Full, stricter \\
\midrule
\input{tab-statements.tex} 
\bottomrule
\end{tabularx}
\end{table}

The statements answer Q4 and Q6 in full, as the schema provides. Under the rule as written, the numbers of statements that answer Q1, Q3 and Q5 in full are \stOneFull, \stThreeFull{} and \stFiveFull; under the stricter reading they are \stOneFullStrict, \stThreeFullStrict{} and \stFiveFullStrict. For Q5 the difference is \nFiveReason{} statements whose text gives, as a finding about the item, a reason narrower than the stated ground: a photograph rejected for a ``Photo Quality Issue'', a picture rejected under a rule against logos or watermarks of other advertising sites, an item restricted ``because it contains pre-owned products''. The statement that is full under both readings removes a browser extension for requesting two named permissions it does not use. The schema has an attribute for why the content falls under the ground, yet in \nFiveNotWhy{} statements (\nFiveNotWhyStrict{} under the stricter reading) no free-text field said why the item fell under the rule. Of these \nFiveNotWhy, \nFiveNo{} restate the ground or the category, describe the process, or name no rule at all, and \nFiveRuleOnly{} name a narrower rule or provision without saying what in the item met it. Q7 is answered in part by \stSevenPartial{} statements, each through the yes-or-no attribute on illegality; no statement says in its own words what the restriction does not establish. Free text raised one of the four questions the schema leaves open in \nStatementsRaised{} statements, under both readings. One describes the user as ``suspected'' of posting prohibited content, which marks the finding as unsettled (Q2, partial). One tells a host that it can return to short-term stays by updating its registration information, a condition under which the restriction would be lifted (Q8: full under the rule as written, partial under the stricter reading). No statement says whether or how the decision could be re-derived, and none names a route for challenge. The keyword screen covers Q2, Q7, Q8 and Q10, \nScreenCells{} cells once grades that rest on a structured field are left out. It agrees with the coding in all but \nScreenDisagree{}: it misses the word ``suspected'', and it counts a reference to ``complaints'' that describes the material considered, not a route of appeal.

Some observations fall outside the protocol. They are examples, not counts, and the list is not exhaustive. One statement says that the date of the content was not known to the team that decided, and that it was copied from the date of the decision so that the statement could be uploaded: the record reports that one of its own fields is not true. One carries the word ``VIDE'' (French for ``empty'') in the required field for the facts relied on, which a check for empty fields accepts. One calls a terminated account suspended and cites a violation report although it gives the platform's own initiative as the source. One lists five possible reasons, of which its keyword field points to the content not being written in German, while it records German as the content's language. Two give a content date one day later than another date in the same statement, the time the statement was received in one and the date the decision applied in the other; since these dates carry no time zone, both may be effects of time zones.

The sample is small, \nStatementsSampleDay{} of its statements come from one day, and it was coded once\ifcodingchecked, by a language model, and checked by the author\else, by a language model, and not yet checked by the author\fi; the grades and the text each rests on are in the ancillary files. Within those limits it adds two things to the schema mapping. The free text does not fill the format's gaps: why a matter was unsettled, what a restriction depends on, whether it can be re-derived and how to challenge it go almost entirely unanswered. And a field the schema provides is not a field that is used: the schema provides an attribute for why the content falls under the ground, a full answer on paper, yet no free-text field said why the item fell under the rule in \nFiveNotWhy{} statements under the rule as written and in \nFiveNotWhyStrict{} under the stricter reading. The expectation recorded in the protocol, that Q5 would mostly be answered in full, was wrong under both readings.

\section{Discussion}\label{sec:discussion}

\emph{What the results show.} Schema validity is weak evidence that a record is adequate. On our records it notices that a required property is missing and nothing else, and it accepts a record that says almost nothing. The four checks together detect every defect we seeded, which shows only that the checks do what they were written to do. The limits lie elsewhere. Template text and a rule applied against its own evidence can be found by a reader but not by a generic check, and falsification cannot be detected from the record at all.

\emph{What sufficiency is for.} The misapplied-rule variant shows the value of the property without overstating it. A sufficient record does not make a decision right. It makes a wrong decision findable by someone who was not present, by re-deriving it, and it tells them what to attack and where. That is the function administrative law assigns to the record and the statement of reasons (Section~\ref{sec:law}).

\emph{When such a record is warranted.} Not every record needs to be a decision record. A timestamped event, a file hash, a model card, a certificate of completion and a model output that nobody adopted are not dispositions of unsettled claims. The test is Q2: if nothing was unsettled, there is nothing to record beyond the event. The format is warranted where a status will be relied on by people who were not party to the decision, and where reading it too broadly would matter.

\emph{Costs and manipulation.} Decision records take time to write and expose the organisation that writes them. Chappidi et al.\ report both that most practitioners who saw record-keeping change their organisation described the change as positive and that some described records being adjusted or falsified, in one account changed in bulk before an audit and changed back afterwards \citep{chappidi2025capture}. A format that asks for more explicit statements gives more to check and more to falsify. The real statements in Section~\ref{sec:statements} show the weak point in small: a required field filled with a copied value so that a statement could be uploaded, and a required field holding the word for ``empty''. Requiring a field produces an entry, not an answer. Two measures follow from the separation in Section~\ref{sec:cfm}. Fidelity should be tested against records the decider does not write, as Zhao et al.\ propose for agents \citep{zhao2026auditable}. And adequacy needs readers, although template text can be flagged mechanically across a set of records, as we did for the shipped examples, though not within one.

\emph{Private evidence.} Some evidence cannot be disclosed. The DSA database shows one resolution: a full statement to the affected person and a reduced public record. Which of the ten questions a public record must still answer for the public's uses is open. In the database we examined, the route of challenge (Q10) and the conditions on which a restriction depends (Q8) are lost, and limits survive only as a yes-or-no field.

\emph{Decisions involving AI.} Nothing in the definition is specific to AI. Where a model contributes, the format keeps its output among the evidence and requires a separate statement of who or what adopted it and under which rule; in the label case the prelabel is recorded and not adopted. The evidence vocabulary of DRK v0.1 has no value for a model output, which we recorded as a system event; a later version should add one.

\section{Limitations}\label{sec:limitations}

\begin{itemize}
\item The case records are synthetic and written by the author of the format. They show that the format can express three kinds of decision. They are not evidence about real records, real organisations or real reviewers.
\item No user study was carried out. We do not know whether readers of a sufficient record can in fact explain, bound, re-derive and challenge a decision better than readers of a conventional record. The four uses are argued for, not measured. Adequacy, which needs a reader, was judged by us for the synthetic records and, for the statements, by the language-model coding\ifcodingchecked{} that the author checked\else, which the author has not yet checked\fi; no independent reader judged either.
\item The ten questions are one decomposition, formulated with DRK v0.1 in view. The argument that each is necessary is a removal test, not a proof; sufficiency of the ten for the four uses is not proved.
\item Checks C2 to C4 and the mutation classes come from the same author, and classes B to D were designed to remain valid under the schema, so every detection pattern in Table~\ref{tab:mutation} follows from the definitions.
\item The mapping of the DSA schema was made by one person, and the coding of the \nStatements{} statements by a language model\ifcodingchecked{} and checked by the author\else{}; the author has not yet checked the grades\fi. Neither was blind, and neither has been repeated by an independent coder. Whether a text is generic is a judgment, and one rule for it was settled during coding.
\item The \nStatements{} statements come from \nStatementPlatforms{} platforms, \nStatementsSampleDay{} of them from one day. They are not a sample of the more than 4.7 billion statements the database listed on 16 November 2026 \citep{ec2026dsadb}, and they are the public database entries, not the statements as sent to recipients.
\item The legal sources motivate the design. No claim is made that a sufficient record satisfies a duty to give reasons, the Digital Services Act, the Artificial Intelligence Act or any other legal requirement.
\item DRK v0.1 is a reference artefact, not a standard. Its vocabularies are incomplete, its schema does not enforce the content the definition requires, and its shipped examples are template text.
\item Redaction, retention, access control and the security of records are not addressed.
\end{itemize}

\section{Conclusion}

Records of decisions commonly preserve what went in and what came out. Record sufficiency names what else a record must preserve so that someone who was not there can explain the status, see its limits, repeat the decision and contest it. Ten questions follow from those four uses. A machine can check that a record answers them in form. Whether the answers are specific enough to use needs a reader, whether they are true needs something outside the record, and sufficiency does not require that the decision was right. On the records examined here, the gap between a valid record and a record that answers the questions is large: a schema validator accepts a record that says almost nothing. On the one statutory format we examined, the schema leaves four questions unanswered (why the matter was unsettled, what the status depends on, whether it can be re-derived, how to challenge it), and in \nStatements{} real statements the free text almost never answers them.

The next steps are empirical. The format should be tried on real decisions by people other than its author, the coding of real records should be repeated by independent human coders, the four uses should be tested with readers, and the question of which answers a public record must keep should be settled case by case.

\section*{Availability}

The ancillary files of this paper contain the DRK v0.1 schema, profiles and examples; the case records and variants and the script that writes them; the four checks and their self-tests; the mutation generator; the DSA attribute list and mapping; the sampling protocol, a saved copy of the database's search page, the \nStatements{} statements as retrieved, the script that holds the coding, and the scripts that drew, check and tabulate the sample; and the results. One command, \code{reproduce.sh}, regenerates every number that comes from the evaluation and the bodies of Tables~\ref{tab:mutation} to~\ref{tab:statements}, checks that the sample follows its seeded order, and fails if any output differs from the version compiled into this paper. It needs Python 3 and the \code{jsonschema} package, and uses no network access and no randomness. The statements of reasons are reproduced from the European Commission's DSA Transparency Database with attribution.

\section*{Use of AI tools}

The research problem, the core conceptual framework, the analytical direction and the conclusions of this paper were defined and determined by the author, building on the author's earlier work \citep{brandes2026trialdiff,brandes2026state} and on DRK v0.1 (Section~\ref{sec:drk}). Anthropic Claude and OpenAI ChatGPT were used to assist with literature discovery and source verification, drafting support and revision, implementation of the evaluation scripts, and critical review of drafts. For the empirical analysis described in Section~\ref{sec:method}, a language model applied the coding protocol to the \nStatements{} statements of reasons; separate language-model agents reviewed the coding, which was revised once in response, \ifcodingchecked and the author then checked and confirmed the grades of all \nStatements{} statements\else and the author has not yet checked the grades\fi. AI-generated draft material and code were evaluated and selectively incorporated by the author, who retains full responsibility for the manuscript, references and reported results.

\section*{Competing interests and licence}

The author has no competing interests. Two earlier works by the author are cited where this paper builds on them \citep{brandes2026trialdiff,brandes2026state}. \textcopyright{} 2026 Amadeus Brandes. This work is licensed under the Creative Commons Attribution 4.0 International licence (CC BY 4.0).

\bibliographystyle{plainnat}
\bibliography{references}

\appendix

\section{Check definitions}\label{app:checks}

\emph{C1 Structure.} The record validates against the DRK v0.1 JSON Schema (draft 2020-12). Values with the \code{date-time} format must be RFC 3339 date-times.

\begin{sloppypar}\emph{C2 Profile.} A profile definition exists for \code{domain\_profile.profile\_id}; \code{domain\_profile.claim\_class} and \code{claim.claim\_class} equal the profile's claim class; every field listed in the profile's \code{profile\_extensions\_required} is present in \code{profile\_extensions} and not empty.\end{sloppypar}

\emph{C3 Content.} The fields in Table~\ref{tab:c3} are not empty. A value is empty if it is null, a string of white space, an empty list or object, or a list of empty values. In addition: if \code{replay.replay\_possible} is true, \code{replay.decision\_inputs} and \code{replay.replay\_instructions} are not empty; if it is false, \code{replay.known\_non\_replayable\_elements} is not empty.

\begin{table}[h]
\caption{Fields checked by C3.}\label{tab:c3}
\footnotesize
\begin{tabularx}{\linewidth}{@{}lL@{}}
\toprule
Question & Fields \\
\midrule
Q1 & \code{claim.statement}, \code{claim.claimant.name}, \code{claim.claimed\_state} \\
Q2 & \code{instability.reason} \\
Q3 & \code{summary} and \code{source} of every item in \code{evidence\_inputs} \\
Q4 & \code{authority\_or\_rule.authority\_name}, \code{disposition.issued\_by} \\
Q5 & \code{authority\_or\_rule.rule\_summary}, \code{authority\_or\_rule.decision\_method}, \code{disposition.rationale} \\
Q6 & \code{disposition.disposition\_value}, \code{disposition.effective\_scope}, \code{horizon\_or\_trigger.description} \\
Q7 & \code{limitations.claims\_not\_supported} \\
Q8 & \code{support\_relation.summary}, \code{support\_relation.failure\_mode} \\
Q9 & the replay rule stated above \\
Q10 & \code{contestability.challenge\_conditions}, \code{contestability.review\_path} \\
\bottomrule
\end{tabularx}
\end{table}

\emph{C4 Coherence.} The following relations hold.
\begin{description}[leftmargin=2.6em,labelwidth=2.0em,style=sameline]
\item[R1] The \code{evidence\_id} values in \code{evidence\_inputs} are unique.
\item[R2] Every entry of \code{authority\_or\_rule.decision\_inputs\_required} is the identifier of a listed evidence input.
\item[R3] Every entry of \code{replay.decision\_inputs} is the identifier of a listed evidence input.
\item[R4] \code{disposition.issued\_by} equals \code{authority\_or\_rule.authority\_name}.
\item[R5] No evidence input has an \code{observed\_at} later than \code{disposition.issued\_at}.
\item[R6] \code{claim.asserted\_at}, if given, is not later than \code{disposition.issued\_at}.
\item[R7] \code{horizon\_or\_trigger.effective\_until}, if given, is not earlier than \code{effective\_from}.
\item[R8] \code{created\_at} is not earlier than \code{disposition.issued\_at} and not more than a set interval after it. The interval is a parameter; the default is 24 hours.
\item[R9] The record does not name itself as its own predecessor or as a record it supersedes.
\end{description}

\end{document}

%% file: results-macros.tex
\newcommand{\nMutants}{164}
\newcommand{\nMutCone}{51}

\newcommand{\nClassA}{51}
\newcommand{\nClassB}{60}
\newcommand{\nClassC}{23}
\newcommand{\nClassD}{30}
\newcommand{\nMissedByCone}{113}
\newcommand{\nBcaughtByCfour}{6}
\newcommand{\nContentFields}{19}
\newcommand{\nContentEmpty}{19}
\newcommand{\nBlankLeaves}{25}
\newcommand{\nStringLeaves}{36}
\newcommand{\nFalseDiff}{15}
\newcommand{\nMisappliedDiff}{5}
\newcommand{\nSorAttr}{35}

\newcommand{\nSorFull}{4}
\newcommand{\nSorPartial}{2}
\newcommand{\nSorNone}{4}
\newcommand{\nSorMissingSections}{4}
\newcommand{\nSorContentFindings}{9}

\newcommand{\nContestLeaves}{8}
\newcommand{\nContestIdentical}{8}

\newcommand{\nExamples}{5}
\newcommand{\nRequiredProps}{17}
\newcommand{\nInstabilityTypes}{12}
\newcommand{\nAuthorityTypes}{11}
\newcommand{\nHorizonTypes}{10}
\newcommand{\nEvidenceTypes}{17}
\newcommand{\nProfiles}{5}
\newcommand{\nProfileExtMin}{7}
\newcommand{\nProfileExtMax}{8}
\newcommand{\nSelfTests}{24}
\newcommand{\nStatements}{25}
\newcommand{\nStatementPlatforms}{25}
\newcommand{\nStatementFallback}{2}
\newcommand{\nPlatformsListed}{396}
\newcommand{\nPlatformsTried}{54}
\newcommand{\nStatementUnavailable}{1}
\newcommand{\nDayQueryFailed}{3}
\newcommand{\nCodingReviewed}{25}
\newcommand{\nStatementsSampleDay}{23}
\newcommand{\fallbackDays}{10 and 13 November}
\newcommand{\nFiveReason}{6}
\newcommand{\nFiveRuleOnly}{3}
\newcommand{\nFiveNo}{15}

\newcommand{\nFiveNotWhy}{18}
\newcommand{\nFiveNotWhyStrict}{24}
\newcommand{\nStatementsRaised}{2}
\newcommand{\nScreenCells}{95}
\newcommand{\nScreenDisagree}{2}
\newcommand{\stOneFull}{7}

\newcommand{\stOneFullStrict}{1}

\newcommand{\stThreeFull}{7}

\newcommand{\stThreeFullStrict}{1}

\newcommand{\stFiveFull}{7}

\newcommand{\stFiveFullStrict}{1}

\newcommand{\stSevenPartial}{5}
\newcommand{\stSevenNone}{20}

%% file: tab-mutation.tex
A\quad Delete one required top-level property & 51 & 51 & 9 & 30 & 9 & 51 \\
B\quad Empty one field that answers a question & 60 & 0 & 0 & 60 & 6 & 60 \\
C\quad Delete one profile-required field & 23 & 0 & 23 & 0 & 0 & 23 \\
D\quad Break one cross-field relation & 30 & 0 & 0 & 0 & 30 & 30 \\
\midrule
All mutants & 164 & 51 & 32 & 90 & 45 & 164 \\

%% file: tab-dsa.tex
Q1 What was asserted, and by whom? & partial & \texttt{puid}, \texttt{content\_type}, \texttt{content\_type\_other} and 10 more \\
Q2 Why was the claim not already settled? & none & none \\
Q3 On what material was it decided? & full & \texttt{decision\_facts}, \texttt{automated\_detection} \\
Q4 Who or what decided? & full & \texttt{platform\_name}, \texttt{automated\_decision} \\
Q5 Under which rule, applied how? & full & \texttt{decision\_ground}, \texttt{decision\_ground\_reference\_url}, \texttt{illegal\_content\_legal\_ground} and 4 more \\
Q6 What status resulted; where and how long does it hold? & full & \texttt{decision\_visibility}, \texttt{decision\_visibility\_other}, \texttt{decision\_monetary} and 9 more \\
Q7 What does the status not establish? & partial & \texttt{incompatible\_content\_illegal} \\
Q8 What must stay true for the status to hold? & none & none \\
Q9 Can the decision be re-derived, and what cannot be? & none & none \\
Q10 On what grounds and where can it be challenged? & none & none \\

%% file: tab-statements.tex
Q1 What was asserted, and by whom? & partial & 7 & 18 & 0 & 1 \\
Q2 Why was the claim not already settled? & none & 0 & 1 & 24 & 0 \\
Q3 On what material was it decided? & full & 7 & 18 & 0 & 1 \\
Q4 Who or what decided? & full & 25 & 0 & 0 & 25 \\
Q5 Under which rule, applied how? & full & 7 & 18 & 0 & 1 \\
Q6 What status resulted; where and how long does it hold? & full & 25 & 0 & 0 & 25 \\
Q7 What does the status not establish? & partial & 0 & 5 & 20 & 0 \\
Q8 What must stay true for the status to hold? & none & 1 & 0 & 24 & 0 \\
Q9 Can the decision be re-derived, and what cannot be? & none & 0 & 0 & 25 & 0 \\
Q10 On what grounds and where can it be challenged? & none & 0 & 0 & 25 & 0 \\